# Heterodyne coherent detection of the electric field temporal trace emitted by frequency-modulated comb lasers

Baptiste Chomet,[1] Salim Basceken,[1] Djamal Gacemi,[1] Barbara Schneider,[2] Mathias Beck,[2] Angela Vasanelli,[1] Benoit Darquié,[3] Jérôme Faist,[2] and Carlo Sirtori[1,*]

[1]*Laboratoire de Physique de l'Ecole normale supérieure, ENS, Université PSL, CNRS, Sorbonne Université, Université Paris Cité, Paris, France*
[2]*Institute for Quantum Electronics, ETH Zürich, Zürich, Switzerland*
[3]*Laboratoire de Physique des Lasers, CNRS, Université Sorbonne Paris Nord, 93430 Villetaneuse, France*
*\*carlo.sirtori@ens.fr*

Frequency-modulated (FM) combs are produced by mode-locked lasers in which the electric field has a linearly chirped frequency and nearly constant amplitude. This regime of operation occurs naturally in certain laser systems and constitutes a valuable alternative to generate spectra with equidistant modes. Here, we use a low-noise fs-pulse comb as the local oscillator and combine dual comb heterodyne detection with time domain analysis of the multi-heterodyne signal to reveal the temporal trace of both amplitude and phase quadratures of FM comb lasers' electric field. This technique is applied to both a dense and a harmonic mid-infrared free-running quantum cascade laser frequency comb and shows direct evidence of the FM behavior together with the high degree of coherence of these sources. Our results furnish a deeper insight on the origin of the FM combs and pave the way to further improvement and optimization of these devices.

## 1. INTRODUCTION

Frequency-modulated (FM) combs are a type of frequency comb that have been observed in different semiconductor lasers: quantum cascade lasers (QCLs) [1–3], quantum dot lasers [4], quantum dash lasers [5], and strained quantum well lasers [6]. All these devices operate in a regime where no pulses are generated, yet they spontaneously produce optical combs in the frequency domain. In terms of laser parameters, they are characterized by the product $f_{rep}\tau_e$ satisfying $10^{-2} < f_{rep}\tau_e < 10$, where $f_{rep}$ is the frequency distance between two longitudinal modes and $\tau_e$ the electronic upper state lifetime [7]. FM combs generate their own frequency modulation, without the need of any additional nonlinear element within the cavity. Their temporal operation is therefore fundamentally distinct from that of conventional mode-locked lasers that generate pulses and operate in the opposite limit, i.e., $f_{rep}\tau_e \gg 1$.

In the last decade different approaches have been developed with the goal either to measure directly or to reconstruct the electric field profile of FM combs. The shifted wave interference Fourier transform spectroscopy (SWIFTS) [8] provides a direct measurement of the phase differences of the adjacent comb modes. The time domain reconstruction relies on performing cumulative sums of spectral phase differences and measurement noise makes it often difficult to retrieve the absolute phases using this technique.

These limitations are particularly pronounced for frequency combs featuring spectral holes in the emission spectrum [9]. Moreover, this technique requires to access the intermodal beating frequency $f_{rep}$, which can be challenging or impossible for high-rate sources such as harmonic combs [10,11]. Dual comb detection schemes allow direct access to absolute spectral phases and amplitudes but require a reference frequency comb of known complex spectrum [12,13]. The temporal field profile of the comb of interest can be reconstructed from the recorded multi-heterodyne signal in the frequency domain considering phase locking between the two combs [14].

Apart from aforementioned methods that rely on frequency domain measurements, optical sampling techniques using a mode-locked femtosecond laser permit to measure light fields directly in time domain. This approach has been notably used for the characterization of semiconductor and Kerr combs around 1.5 μm [15,16] and has proven particularly successful in the terahertz frequency range using the electro-optic effect in a ZnTe crystal [17]. Recently such a technique has been used in the mid-infrared (MIR) region using a sum-frequency generation process, giving access to the down sampled laser *intensity* only [18].

In the MIR, QCLs have emerged as the dominant source of coherent light of particular interest for spectroscopic applications

[19,20]. They exhibit homogeneously broadened transitions and spatial hole burning effects, allowing multimode operation and self-generation of frequency combs for a certain amount of intra-cavity phase dispersion [21,22]. In addition, the ultra-short relaxation time of the carriers, $\tau_e \approx 1$ ps, leads to an efficient four wave mixing (FWM) process ($f_{rep}\tau_e \ll 1$) that tends to minimize carrier fluctuations and maximize the mean extracted power [23]. Then QCL combs typically operate in FM regime with near constant intensity but produce a linear frequency chirp that spans the whole laser spectrum [2,8]. It can also manifest as harmonic states, where the FM behavior repeats an integer number of times per round trip [7,10].

In this paper we demonstrate for the first time to our knowledge optical sampling directly in the MIR of the electric field profile emitted from free-running QC comb lasers. By using a stabilized low-noise commercial mode-locked femtosecond laser emitting directly in the MIR as the local oscillator (LO), we are able to recover, on a 50 MHz bandwidth detector, the slowly varying replica of the temporal waveform of QC lasers. With this setup we have investigated the electric field emitted from two comb QCLs: a fundamental mode-locked laser and a harmonic mode-locked laser. For both devices we were able to precisely measure the time evolution of phase and amplitude in excellent agreement with theoretical predictions. Moreover by using our LO comb as an optical reference, we demonstrate active stabilization of QCL combs and we show that this setup allows to characterize the coherence throughout its entire spectrum. Our work is thus an important step toward the development of powerful and ultra-stable QCL frequency combs for both fundamental research and technological applications.

## 2. METHODS

The basics of our detection technique is illustrated in Fig. 1(a). The signal under test, a QCL comb source, produces an unknown signal whose intensity $I_{QCL}(t) = |E_{QCL}(t)|^2$ and phase $\phi_{QCL}(t)$ are periodic:

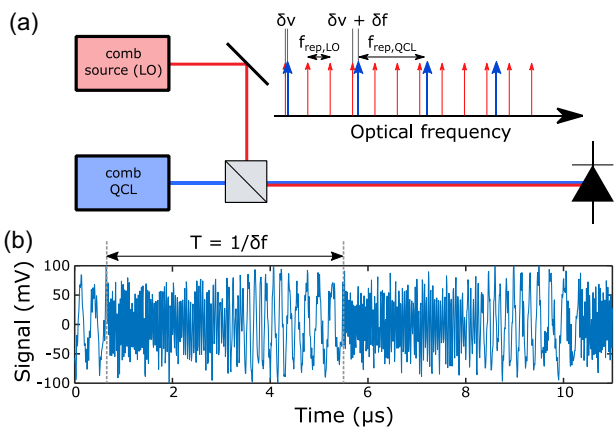

**Fig. 1.** (a) Conceptual scheme of the multi-heterodyne detection that uses two combs with a very different repetition rate. (b) Measured signal $S(t)$ as a function of time over two "frames".

$$E_{QCL}(t) = e^{i2\pi f_{0,QCL}t}\sqrt{I_{QCL}(t)}e^{i\phi_{QCL}(t)}$$
$$= e^{i2\pi f_{0,QCL}t}\sum_{n=-N_{QCL}/2}^{n=N_{QCL}/2} A_{QCL,n}e^{i2\pi nt f_{rep,QCL}}e^{i\varphi_{QCL,n}}, \quad (1)$$

with $f_{0,QCL}$ the frequency of the optical carrier, and $N_{QCL}$ the number of oscillating modes with amplitude $A_{QCL,n}$ and phase $\varphi_{QCL,n}$ (with $n$ an integer) and mode spacing $f_{rep,QCL}$. The LO comb has the same form, with the subscript QCL → LO. When the two combs are mixed the detected heterodyne signal is $S(t) = R_{det}(t) * \text{Re}[E_{QCL}(t)E_{LO}^*(t)]$, where $R_{det}(t)$ is the response of the detector convoluted with that of any frequency filter used in data acquisition. The symbol $*$ represents a convolution product. From the frequency domain picture in Fig. 1(a), it can be readily seen that radiofrequency (RF) beat-notes will appear at each frequency difference between pairs of comb lines. The effective detuning between the repetition rates, $\delta f$, determines the beat-note spacing in the RF domain. It should be noted that for large detunings or a large interacting bandwidth, some beat-notes may get to the point where the next closest comb line is not to the lower- but to the higher-frequency side. When this occurs, a different set of beat-notes with the same spacing but at a different offset will appear within the band of interest, which generates aliasing. To avoid this, the following condition must be satisfied:

$$N|\delta f| + \delta \nu = N|f_{rep,QCL} - k f_{rep,LO}| + \delta \nu < \frac{f_{rep,LO}}{2}. \quad (2)$$

In this expression, $N$ is the number of interacting comb lines, $\delta f = f_{rep,QCL} - k f_{rep,LO}$ is the effective repetition rate detuning, and $\delta \nu$ is the frequency difference between the two closest QCL and LO-comb optical lines. In particular, when the LO comb produces short pulses, modal phases are constant in time and their phase relation is linear ($\varphi_{LO,n} = n \times \theta$ with $\theta$ a constant phase difference), while the amplitude is constant along the QCL-comb emitted spectrum. As a consequence, the complete time evolution of the QCL field including its phase can straightforwardly be extracted from the recorded multi-heterodyne beat signal. By limiting the responsivity bandwidth $R_{det}(t)$ using a bandpass filter with cut-off frequency $< f_{rep,LO}/2$, the heterodyne signal becomes

$$S(t) = R_{det}A_{LO}\text{Re}\left[e^{i\phi_0}e^{i2\pi \Delta f_0 t}\sum_{n=-\frac{N}{2}}^{n=\frac{N}{2}} A_{QCL,n}e^{i2\pi nt\delta f}e^{i\varphi_{QCL,n}}\right]$$
$$= R_{det}A_{LO}\text{Re}\left[e^{i\phi_0}e^{i2\pi \Delta f_0 t}\sqrt{I_{QCL}^{\delta f}(t)}e^{i\phi_{QCL}^{\delta f}(t)}\right]. \quad (3)$$

In this equation, $I_{QCL}^{\delta f}$ and $\phi_{QCL}^{\delta f}$ are a slowly varying replica of the original QCL intensity $I_{QCL}$ and phase $\phi_{QCL}$ of Eq. (1) in which the repetition rate $f_{rep,QCL}$ has been replaced by $\delta f$. The signal $S(t)$ oscillates at the center frequency $\Delta f_0 = \delta \nu + \delta f N/2$ and the sum reflects the repetitive nature of the measurement with repetition rate $\delta f$. The phase offset $\phi_0 = k\theta \Delta f_0/\delta f$ is of subordinate physical relevance, as it only introduces a constant time delay in the recorded waveform. $R_{det}$ represents the detector response that is spectrally flat in the RF 50 MHz band and whose optical band is very large compared to our comb bandwidth.

The laser used to obtain the signal shown in Fig. 1(b) is a QCL emitting at $\lambda = 8.1$ µm with a mode spacing $f_{rep,QCL} = 7.4$ GHz.

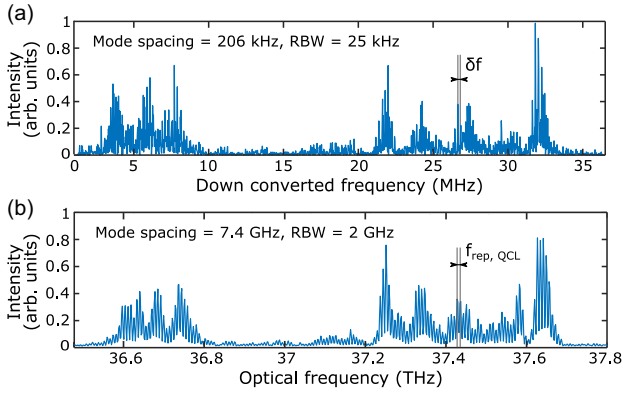

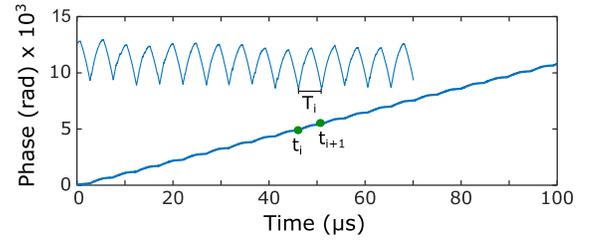

**Fig. 2.** (a) Radiofrequency spectrum obtained by the Fourier transform of the down converted temporal trace $S(t)$ on a sample of 40 µs. (b) Optical spectrum of the source under test recorded with an optical spectrum analyzer of 2 GHz resolution bandwidth (RBW).

**Fig. 3.** Measured unwrapped phase, $2\pi \Delta f_0 t + \phi_{\text{QCL}}^{\delta f}(t)$, of the signal $S(t)$ [see Eq. (3)] of the fundamentally mode-locked laser as a function of time. In the top left part, the periodic contribution to the phase is also shown with a zoom ×25. This is obtained by subtracting the contribution of the center frequency $\Delta f_0$ (17.29 MHz here) from the unwrapped phase.

The local oscillator is a turn-key offset-free mid-IR mode-locked laser in the short pulse regime from Menlo Systems spanning the 7.4 to 9.4 µm spectral window (full-width-half-maximum FWHM ∼ 4.69 THz) with mode spacing $f_{\text{rep,LO}} = 100$ MHz. In experiments, it is possible to fulfill Eq. (2) by tuning the repetition rate of the QCL with current (the corresponding tuning rate is −100 kHz/mA) to be close to the $k = \text{round}[f_{\text{rep,QCL}}/f_{\text{rep,LO}}] = 74^{\text{th}}$ harmonics of the LO-comb repetition rate. In addition, the QCL optical line can be tuned (the corresponding tuning rate is −235 MHz/mA) with current for $\delta \nu$ to be close to the zero frequency. The signal $S(t)$ at the output of a mercury-cadmium-telluride (MCT) detector (VIGO System PVI-4TE-10.6) is then recorded in the temporal domain on a fast oscilloscope with a sampling rate of 250 MS/s after a low-pass filter of 48 MHz. The resulting waveform given by Eq. (3) then gives a snapshot of the QCL signal, including the carrier under the periodic envelope with periodicity $T = 1/\delta f$. Figure 1(b) shows the signal $S(t)$ as recorded in a real time oscilloscope over 10 µs, which contains two frames of length $T \sim 5$ µs, corresponding to a copy of the optical signal temporally stretched by a factor $M = f_{\text{rep,QCL}}/\delta f \sim 35920$. The envelope of the signal is almost constant and the carrier exhibits a frequency chirp as expected for FM combs.

Fourier analysis of the time trace recorded over 40 µs is shown in Fig. 2(a), showing that the down converted RF spectrum reproduces with a very good fidelity the one recorded in the optical domain [Fig. 2(b)]. The repetition rate detuning is here $\delta f = 206$ kHz and can be farther decreased to generate waveforms with a longer periodicity in the RF domain. The detuning can be reduced until the periodicity exceeds the coherence length of the source. In our measurements, $T = 1/\delta f$ is approximately 5 µs, and for this integration time the QCL linewidth is Fourier limited, thus enabling coherent detection (see Supplement 1).

## 3. RESULTS

In this section, we apply our coherent sampling technique to retrieve the emission profile of two QCL devices with different frequency separation between the modes. Both devices have the same active region, made of strain-compensated dual-stack heterostructure grown by molecular beam epitaxy. They are cooled to 10°C using a Peltier element and emit more than 100 mW in continuous wave operation. The first QCL investigated has a 6 mm long ridge and is fundamentally mode locked (FML) where $f_{\text{rep}} = 7.4$ GHz corresponds to the free spectral range (FSR) of the cavity. The second one has a 2.9 mm long ridge and operates in the harmonic mode-locked (HML) frequency comb regime [10], where now $f_{\text{rep}} = 60$ GHz $= 4 \times$ FSR, with FSR $= 15$ GHz. For the latter, no beating at the cavity FSR frequency was observed in the RF spectrum of the laser intensity.

The numerical acquisition of the time domain field profile is the central entry point. From this, as Eq. (3) shows, it is possible to extract the center frequency contribution $\Delta f_0$ as well as the periodic field $\xi_{\text{QCL}}(t) = \sqrt{I_{\text{QCL}}^{\delta f}(t)} e^{i\phi_{\text{QCL}}^{\delta f}(t)}$ in the time domain thanks to the numerical Hilbert transform H. Figure 3 shows the unwrapped phase of the heterodyne beating of Fig. 1(b), which can be written as the sum of a contribution from the heterodyne center frequency $2\pi \Delta f_0 t$ and a periodic phase $\phi_{\text{QCL}}^{\delta f}$ with time periodicity $T$. When the laser operates in the free-running regime, due to cavity fluctuations, the time between two successive frames is not constant and therefore we can define $T_i = t_{i+1} - t_i$, where $t_i$ is the starting time of the $i$th frame. Each $T_i$ can be assessed directly through analysis of the periodic fluctuations of the signal intensity (see Supplement 1). From this, an averaging over 1000 frames of the phase $\phi_{\text{QCL}}^{\delta f}(t)$ and the intensity of the periodic field $\xi_{\text{QCL}}(t)$ is illustrated in Fig. 4 over two round-trip times of the cavity $T_{\text{cav},M}$ in the laboratory time scale, i.e., the cavity round-trip time $T_{\text{cav}} = 1/\text{FSR}$ magnified by the factor $M$.

We begin by investigating the QCL operating in the FML regime; phase and intensity of its temporal waveform are respectively shown in Figs. 4(a) and 4(b). In this regime, $f_{\text{rep}}\tau_e \sim 0.01 \ll 1$, and a parabolic behavior of the phase is observed together with a nearly constant intensity. This shows that the laser always tends to suppress intensity fluctuations despite its spectral mode distribution. We find the results in excellent agreement with theoretical modelling [22], which predicts the phase curvature assuming constant amplitude following

$$\frac{\partial^2 \phi_{\text{QCL}}}{\partial \tau^2} = -2\pi \Delta f \times f_{\text{rep}} = -4\pi^2 f_{\text{rep}}^2 \left( \frac{\partial^2 \varphi_{\text{QCL},n}}{\partial n^2} \right)^{-1}, \quad (4)$$

with $\tau$ the time within one period of intermodal beats $T_{\text{rep}} = 1/f_{\text{rep}}$, and $\Delta f \sim 1.2$ THz the spectral bandwidth of the comb emission. The curvature of the parabolic red line in Fig. 4(a) satisfies Eq. (4) and shows that the laser emission frequency is chirped over the whole laser spectrum. Moreover, in the present case of a purely quadratic phase behavior, Eq. (4) dictates a simple relation between the phase measured in the time domain and the spectral phase distribution. We retrieve the intensity and phase of each optical mode [$|A_{QCL,n}|^2$ and $\varphi_{QCL,n}$; see Eq. (3)] by applying Fourier transform over samples of period $T = 1/\delta f$ of the down sampled periodic field $\xi_{QCL}(t)$. Figure 4(e) shows the optical spectrum of the FML QCL with a spectral gap typical of QCL frequency comb emission. The spectral distribution of the modal phases $\varphi_{QCL,n}$ is shown in Fig. 4(f) where the data are compared with the results of Eq. (4) (in red dashed line). The agreement between time and frequency domains related by Eq. (4) is excellent as the FML QCL shows a close to ideal phase modulated waveform.

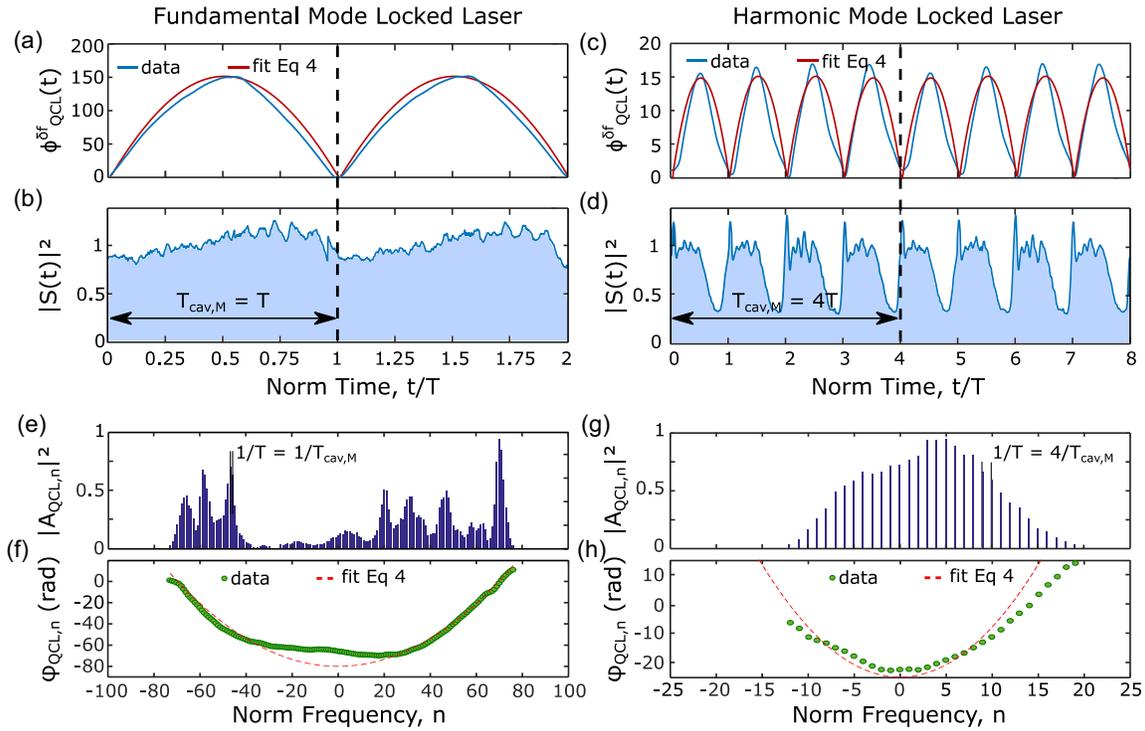

**Fig. 4.** Phase and intensity of the waveform over two cavity round-trips in the laboratory time scale $T_{cav,M} = M \times T_{cav}$ for the (a), (b) FML QCL and (c), (d) HML QCL. The phase is compared with theoretical results from Eq. (4) in red. The time is normalized to the time between two successive frames $T = M \times T_{rep}$. Spectra $|A_{QCL,n}|^2$ and modal phases $\varphi_{QCL,n}$ for the (e), (f) FML QCL and (g), (h) HML QCL. The phase is compared with theoretical results from Eq. (4) in red dashed line. The frequency is normalized to the frequency difference between two modes $1/T$.

We now investigate the temporal waveform emitted by the HML QCL where the product $f_{rep}\tau_e$ has gained an order of magnitude. The temporal phase is shown in Fig. 4(c) and its curvature shows that the laser emission frequency spans the entire comb bandwidth ~1.2 THz within one period $T_{rep} = 17$ ps. In contrast with the emission of the FML QCL, the intensity of the HML device exhibits a modulation depth reaching more than 75%. In fact, the strength of the FWM process depends, in first approximation, on the previously defined product $f_{rep}\tau_e$, where a high-frequency population inversion increases the gain seen by each sideband in a parametric enhancement effect [24]. However, in the HML frequency comb regime the population inversion can no longer follow exactly the intensity modulation. In this situation, the laser behavior cannot be considered anymore as an ideal FM comb, but rather as being in a hybrid situation in which the frequency modulation is also accompanied by residual amplitude modulation. Such a hybrid regime has been observed in many semiconductor lasers, as we mentioned in Section 1, and also recently in interband cascade lasers [25]. For the majority of these devices the ratio $f_{rep}\tau_e$ gets close to unity. Intensity and phase of each optical mode are plotted in Figs. 4(g) and 4(h), respectively. Notice that in Fig. 4(h) the curvature of the spectral phase distribution given by Eq. (4) is less accurate than for the FML device [Fig. 4(g)].

## 4. LASER STABILIZATION AND NOISE CHARACTERIZATION

Despite the fact that free-running QCL combs exhibit typical linewidths around half a MHz [26], they have proven useful in a dual comb scheme for MIR spectroscopy [27,28]. In this configuration, a tight lock of the dual comb signal between two QCL combs was obtained through a combination of RF injection, optical injection [29], or feedback loops acting on the laser current [30] or more recently on the cavity refractive index [31]. However, the full stabilization of a QCL comb could provide a set of narrow-linewidth and absolute frequency references, paving the way towards metrological and high-resolution spectroscopic applications [32]. As a matter of fact, QCL combs emitting in the THz range have already been fully referenced and used for spectroscopy [33]. Full stabilization of MIR QCL combs on a molecular transition has been reported [34], but the stability of only a single comb line was obtained by heterodyne beating with a stabilized DFB laser.

In this section we demonstrate full stabilization of our QCL comb using our LO comb as an optical reference and show that we can characterize its stability in terms of residual phase fluctuations though its entire spectrum. The repetition rate is locked by RF injection and stabilization of the offset frequency is obtained by phase locking a tooth of the QCL comb to a tooth of the LO comb using a beat-note signal generated directly in the mid-IR region [35] as described in Supplement 1. The other teeth of the LO comb are used to characterize the quality of the stabilization of the QCL comb over its entire spectral coverage through multi-heterodyne. Our approach differs from a line by line phase noise analysis since the recorded time trace allows simultaneous detection of the phase noise of each comb line (see Supplement 1).

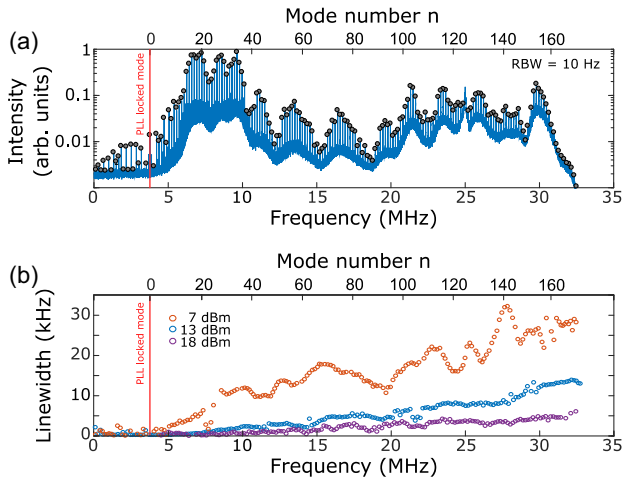

**Fig. 5.** (a) Down converted spectrum in the MHz domain of the fully stabilized QCL comb operating in the FML regime with 10 Hz of resolution. The RF injected power for stabilization of the round-trip frequency is 18 dBm. (b) Linewidth of the stabilized QCL comb modes integrated over 100 ms for different RF injection powers.

The device under investigation is the fundamental mode-locked QCL. The latter does not operate in free running anymore, the optical modes are now stabilized using a combination of RF injection and feedback loops. All details on the stabilization are given in Supplement 1. While the phase difference between one QCL optical mode and its nearest LO-comb tooth is stabilized with a phase-locked loop (PLL), the mode spacing of the QCL comb is stabilized through injection locking, by feeding into the QCL a microwave signal approaching the QCL cavity round-trip frequency. The spectrum obtained by Fourier transform of a 100 ms trace, with 18 dBm of microwave injection power, is shown in log scale in Fig. 5(a). The down converted spectrum spans 35 MHz and the effective spacing between the modes is 170 kHz. To assess the stability of our QCL comb, we computed the Allan deviation of different RF comb lines, which are presented in Supplement 1. From the average of 150 comb lines we obtained a stability of $5 \times 10^{-12}$ at 20 ms; therefore the different optical modes are clearly identified. This is in contrast to a laser operating in the free-running regime, where the different comb teeth overlap due to the broadening of the modes (around 400 kHz $> \delta f$ for 100 ms of integration time; see Supplement 1). The linewidth $\delta \nu_n$ (full-width-half-maximum, 100 ms integration time) of the different optical modes is shown in Fig. 5(b) when the QCL repetition rate is stabilized with different RF power of the synthesizer. The latter is seen to increase linearly with the mode number $n$, where $n = 0$ corresponds to the PLL phase-locked QCL optical line located in the red side of the spectrum. In fact, $\delta \nu_n$ can be written as $\delta \nu_n = \delta \nu_{n=0} + n\beta$, where $\beta$ is the frequency deviation per modes. We confirm that this behavior is due to some residual fluctuations of the cavity by accessing the frequency fluctuations of the repetition rate as described in Supplement 1. As shown in Fig. 5(b), $\beta$ decreases from 190 Hz to 30 Hz with an increase of the injection power from 7 to 18 dBm, meaning that the locking quality is improved with an increase power of the synthesizer. Since the linewidth of the modes is below 10 kHz, it would be possible to stretch the temporal waveform by another factor 10 ($\delta f \sim 20$ kHz) and therefore compress the entire signal in an even narrower frequency band. Moreover, such a metrological control of the comb frequencies opens the path to the generation of exactly periodic electric fields that can be averaged so as to avoid having to record long time traces [29].

## 5. CONCLUSION

In this work, we present a temporal analysis of the electric field emitted by two quantum cascade laser combs operating in the fundamental and harmonic regimes. The experimental technique is based on a multi-heterodyne detection scheme that exploits a low-noise broadband (more than 4 THz) comb as the local oscillator to down convert ultrafast field oscillations, using a 50 MHz detector. Down conversion temporally stretches the optical signal by a factor $5 \times 10^4$ and allows the use of conventional electronics to image the temporal traces of the electric field. From the analysis of the temporal traces it is possible to retrieve a wealth of information that characterizes the laser comb operation. While the dense frequency comb laser reveals a close to ideal frequency-modulated behavior, the harmonic comb shows a strong intensity modulation, similar to that obtained in semiconductor diode lasers. Indeed, due to the very short repetition rate, the harmonic comb is characterized by a product $f_{rep}\tau_e \sim 1$ and the gain cannot follow adiabatically the field in the cavity. These results on the QC laser harmonic comb and their interpretation give clear indications on the mode-locking regime of diode lasers. Finally, there are several applications that could be realized by using this setup: high-resolution time resolved spectroscopy [36,37], phase and amplitude characterization of light modulation [38], and wide-band multi-heterodyne detection of incoherent light sources. Moreover the use of unipolar quantum devices, instead of conventional MCT detectors, can drastically improve the signal to noise ratio and allows applications in quantum sensing [39,40].

**Funding.** ENS-Thales Chair; PEPR Electronique project COMPTERA; CNRS Renatech network; ANR project CORALI (ANR-20-CE04-0006); Ile-de-France Region DIM SIRTEQ.

**Acknowledgment.** We thank Olivier Lopez for providing the stabilization electronics and Marco Piccardo for useful discussions.

**Disclosures.** The authors declare no conflicts of interest.

**Data availability.** Data underlying the results presented in this paper are not publicly available at this time but may be obtained from the authors upon reasonable request.

**Supplemental document.** See Supplement 1 for supporting content.